# Reconfigurable SWCNT ferroelectric field-effect transistor arrays

Dongjoon Rhee,[1,2,7] Kwan-Ho Kim,[1,7] Jeffrey Zheng,[3] Seunguk Song,[1,4,5] Lian-Mao Peng,[6] Roy H. Olsson III,[1] Joohoon Kang,[2,*] and Deep Jariwala[1,*]

[1]Department of Electrical and Systems Engineering, University of Pennsylvania, Philadelphia, PA, USA

[2]School of Advanced Materials Science and Engineering, Sungkyunkwan University (SKKU), Suwon 16419, Republic of Korea

[3]Department of Materials Science and Engineering, University of Pennsylvania, Philadelphia, PA, USA

[4]Department of Energy Science, Sungkyunkwan University (SKKU), Suwon 16419, Republic of Korea

[5]Center for 2D Quantum Heterostructures, Institute for Basic Science (IBS), Sungkyunkwan University (SKKU), Suwon 16419, Republic of Korea

[6]Center for Carbon-Based Electronics, School of Electronics, Peking University, Beijing, China

[7]These authors contributed equally: Dongjoon Rhee, Kwan-Ho Kim
[*]Corresponding author: Deep Jariwala (email: dmj@seas.upenn.edu), Joohoon Kang (email: joohoon@skku.edu)

## Abstract

Reconfigurable devices, capable of switching between multiple circuit elements, have garnered significant attention for alleviating the scaling requirements of conventional complementary metal-oxide-semiconductor (CMOS) technology, as they require fewer components to construct circuits with similar function in traditional CMOS. Prior demonstrations in such devices required continuous voltage application for programming gate terminal(s) in addition to the primary gate terminal, which undermines the advantages of reconfigurable devices in realizing compact and power-efficient integrated circuits. Here, we realize reconfigurable devices based on a single-gate field-effect transistor (FET) architecture by integrating semiconducting channels consisting of a monolayer film of highly aligned single-walled carbon nanotubes (SWCNTs) with a ferroelectric aluminum scandium nitride ($Al_{0.68}Sc_{0.32}N$) gate dielectric, all compatible with CMOS back-end-of-line (BEOL) processing. Leveraging the scalability of both SWCNT assembly and $Al_{0.68}Sc_{0.32}N$ growth processes, we demonstrated these SWCNT ferroelectric FETs (FeFETs) in a centimeter-scale array (~1 cm$^2$) comprising ~735 devices, with high spatial uniformity in device characteristics across the array. The devices exhibited ambipolar transfer characteristics with high on-state currents and current on/off ratios exceeding $10^5$, demonstrating an excellent balance between electron and hole conduction (~270 µA µm$^{-1}$ at a drain voltage of 3 V)—an achievement that has been challenging with previous ambipolar devices based on silicon and other semiconductors. When functioning as a non-volatile memory, the SWCNT FeFETs demonstrated large memory windows of 0.26 V nm$^{-1}$ and 0.08 V nm$^{-1}$ in the hole and electron conduction regions, respectively, combined with excellent retention behavior for up to $10^4$ s. Repeated reconfiguration between *p*-FET and *n*-FET modes was also enabled by switching the spontaneous polarization in $Al_{0.68}Sc_{0.32}N$ and operating the transistor within a voltage range below the coercive voltage. As a proof-of-concept demonstration, we revealed through computer-aided circuit simulations that reconfigurable SWCNT transistors can realize ternary content-addressable memory (TCAM) with far fewer devices compared to circuits based on conventional silicon CMOS technology or alternative systems based on resistive non-volatile devices.

**Introduction**

With the advancement of artificial intelligence (AI) and the Internet of Things (IoT), the demand for processing large volumes of data is rapidly increasing[1,2]. However, scaling both device dimensions and supply voltage to meet this demand are becoming more challenging with conventional silicon (Si) complementary metal-oxide-semiconductor (CMOS) technology[3,4]. In particular, further size reduction of field-effect transistors (FETs) is hindered by various short channel effects[3,4], while reducing the supply voltage to deep sub-1 V levels is constrained by the Boltzmann limit[4,5]. These limitations not only hinder efforts to increase the device density but also result in substantial power dissipation[4].

Reconfigurable devices, which can be configured into various circuit elements post fabrication, have garnered significant attention to address the challenges associated with device downscaling and power consumption [6-13]. Most notable among these devices is a single reconfigurable device which can dynamically switch between *p* and *n* polarity in a FET geometry[6-8]. This reconfigurable aspect allows a circuit to perform multiple functions within a given physical layout, thereby reducing both the number of devices and the overall circuit dimensions needed to achieve specific functions[10,14-17]. For instance, a single circuit consisting of reconfigurable FETs demonstrated the ability to perform various Boolean logic functions, each requiring fewer unit devices compared to individual logic circuits based on standard CMOS transistors[10,14-16]. Additionally, in the design of a full adder, the use of reconfigurable FETs led to a 40% reduction in area and a 25% reduction in critical path delay[15,17,18]. Power consumption and circuit delay can also be reduced by using reconfigurable device-based circuits compared to CMOS counterparts, as fewer devices are required to achieve the same function[14,16,17]. Despite these notable advantages, the practical application of reconfigurable devices over CMOS technology is hindered by their reliance on Si nanowire channels for sufficient electrostatic doping[6,7]. These nanowires are either produced by bottom-up synthesis, which prevents deterministic fabrication for high device integration density[8,13,19], or by top-down methods with complex processes that lead to low throughput[9,11]. Achieving balanced electron and hole current levels in Si channels is also challenging due to the marked differences in electron and hole effective masses[8,13,20,21].

Low-dimensional semiconductors exhibiting ambipolar carrier transport behavior offer a unique advantage over Si, as their atom-scale thickness and dangling-bond-free surface enable effective electrostatic doping of both electrons and holes and a much smaller scaling length compared to bulk semiconductors[22-25]. Among various ambipolar low-dimensional semiconductors, such as two-dimensional (2D) black phosphorous (BP), tungsten diselenide ($WSe_2$), and molybdenum ditelluride ($MoTe_2$), and one-dimensional (1D) single-walled carbon nanotubes (CNTs)[22,26], SWCNTs are particularly promising due to their nearly equal effective masses for electron and hole, which facilitates highly balanced current levels in both *n*- and *p*-type FET operations[26-28]. Additionally, SWCNTs can be produced at large volumes with over 99.9999% semiconducting purity using well-established growth and solution-based sorting processes[29], while are also being compatible with back-end-of-line (BEOL) processing for monolithic integration with silicon CMOS technology[30,31], which further enhances their merit. However, obtaining high-performance, high reliability, and low variation semiconducting channels from SWCNTs is challenging because the SWCNTs must be assembled from solution into films at high density, with their longitudinal direction aligned parallel to the channel[29,32-34].

In addition to the discovery of alternative channel materials, improvements in device architecture are crucial to fully leverage the benefits of reconfigurable devices. Specifically, reconfigurable FETs require additional gate terminal(s) to program the polarity of the channel (program gate), which poses a drawback for device downscaling compared to conventional three-terminal FETs consisting of a source, drain, and gate[6-13]. Moreover, voltage must be continuously applied to the program gate to maintain the designed polarity of the device, which hinders efforts to lower power consumption[6-13]. Incorporating a ferroelectric insulator into the gate stack is a promising approach to reduce both device dimensions and power consumption. The charge carrier polarity and threshold voltage can be controlled through spontaneous polarization in the ferroelectric layer, which remains non-volatile after the programming step[35,36], thus allowing for device reconfiguration without the need for additional gate terminals and continuous gate biasing. Although SWCNT-based ferroelectric FETs (FeFETs) have been demonstrated using ferroelectric metal oxides and polymers as gate insulators, only $p$-type operation has been achieved[37-45]. The limited tunability of charge carrier polarity in previous SWCNT FeFETs arises because water and oxygen adsorbates can screen the electric field under positive gate bias, rendering the ferroelectric polarization insufficient to induce electron doping[46,47]. Ferroelectric materials with high polarizability offer the potential for enabling electrostatic $n$-doping of SWCNT FeFETs. Among various candidate materials, aluminum scandium nitrides ($Al_{1-x}Sc_xN$) have recently attracted significant attention due to their high remnant polarization ($> 100$ $\mu C/cm^2$)[35,48], which is much higher than that of state-of-the-art hafnium zirconium oxide (HZO) films (1–40 $\mu C/cm^2$)[49,50]. Additionally, $Al_{1-x}Sc_xN$ is compatible with CMOS processes and can be deposited at BEOL-compatible growth temperature ($< 400$ °C)[35,36,51], thus serving as an ideal option for integration with SWCNTs to realize BEOL-compatible reconfigurable devices.

Here, we demonstrate reconfigurable SWCNT FeFETs that can switch between $p$- and $n$-type FETs, as well as a non-volatile memory (NVM), by employing ferroelectric $Al_{0.68}Sc_{0.32}N$ film as the gate dielectric. A dense monolayer film of semiconducting SWCNTs was first assembled with a high degree of alignment on an $Al_{0.68}Sc_{0.32}N$-deposited Si wafer to fabricate centimeter-scale device arrays with high on-state current and gate tunability. The scalability of SWCNT and $Al_{0.68}Sc_{0.32}N$ films enabled reliable and spatially uniform device characteristics across the array. By engineering the metal−semiconductor contact interface, we achieved ambipolar behavior in the SWCNT FeFETs, with well-balanced, high on-state hole and electron current densities (~270 µA µm$^{-1}$ at a drain voltage of 3 V) and current on/off ratios exceeding 10$^5$, placing our device among the highest-performing ambipolar FeFETs reported. When operated as an NVM, the SWCNT FeFET exhibited large memory windows of 0.26 V nm$^{-1}$ and 0.08 V nm$^{-1}$ in the hole and electron conduction regions, respectively, with excellent retention behavior up to 10$^4$ s. Reconfiguration between *p*- and *n*-channel FETs was also achieved by first switching the ferroelectric polarization direction and then operating the device within a gate voltage range below the coercive voltage. Both the on-state current and current on/off ratio remain similar and balanced throughout the repeated polarity switching cycles, highlighting the potential of SWCNT FeFETs in reconfigurable circuits. Lastly, we demonstrate ternary content-addressable memory (TCAM) based on SWCNT FeFETs through computer-aided circuit simulation, using significantly fewer devices compared to the conventional TCAMs that rely on multiple silicon FETs or alternative technologies based on NVM devices.

## Results and Discussion

**Figure 1a** schematically illustrates the SWCNT FeFET fabricated in a back-gate FET configuration, which consists of a monolayer film of aligned semiconducting SWCNTs (average diameter: 1.5 nm[29]) as the channel and a 45-nm Al$_{0.68}$Sc$_{0.32}$N (hereafter denoted as AlScN) ferroelectric insulator film as the gate dielectric. A wurtzite-phase AlScN layer was first epitaxially grown on a platinized 300 nm SiO$_2$/silicon (100) wafer, followed by the poly(methyl methacrylate) (PMMA)-assisted wet transfer process of a highly aligned SWCNT film from an SiO$_2$/Si substrate (**Methods** and **Figure S1**)[52,53]. Notably, the dense and highly aligned nature of the monolayer SWCNT film was preserved during the transfer process, as confirmed by atomic force microscope (AFM) images of the film before and after the transfer (**Figure S2**–S**3**). After the SWCNT transfer process, the channel and source/drain (S/D) electrodes were fabricated with the channel length oriented parallel to the SWCNT alignment to achieve high current density, while the AlScN in the corner region of the substrate was etched to expose the Pt back gate (G) (**Methods**). When a positive gate voltage ($V_{GS}$) exceeding the positive coercive voltage ($V_{GS} > V_{C+}$) is applied, the polarization of the AlScN film orients toward the channel, thereby accumulating electrons in the channel. Conversely, applying $V_{GS}$ more negative than the negative coercive voltage ($V_{GS} < V_{C-}$)

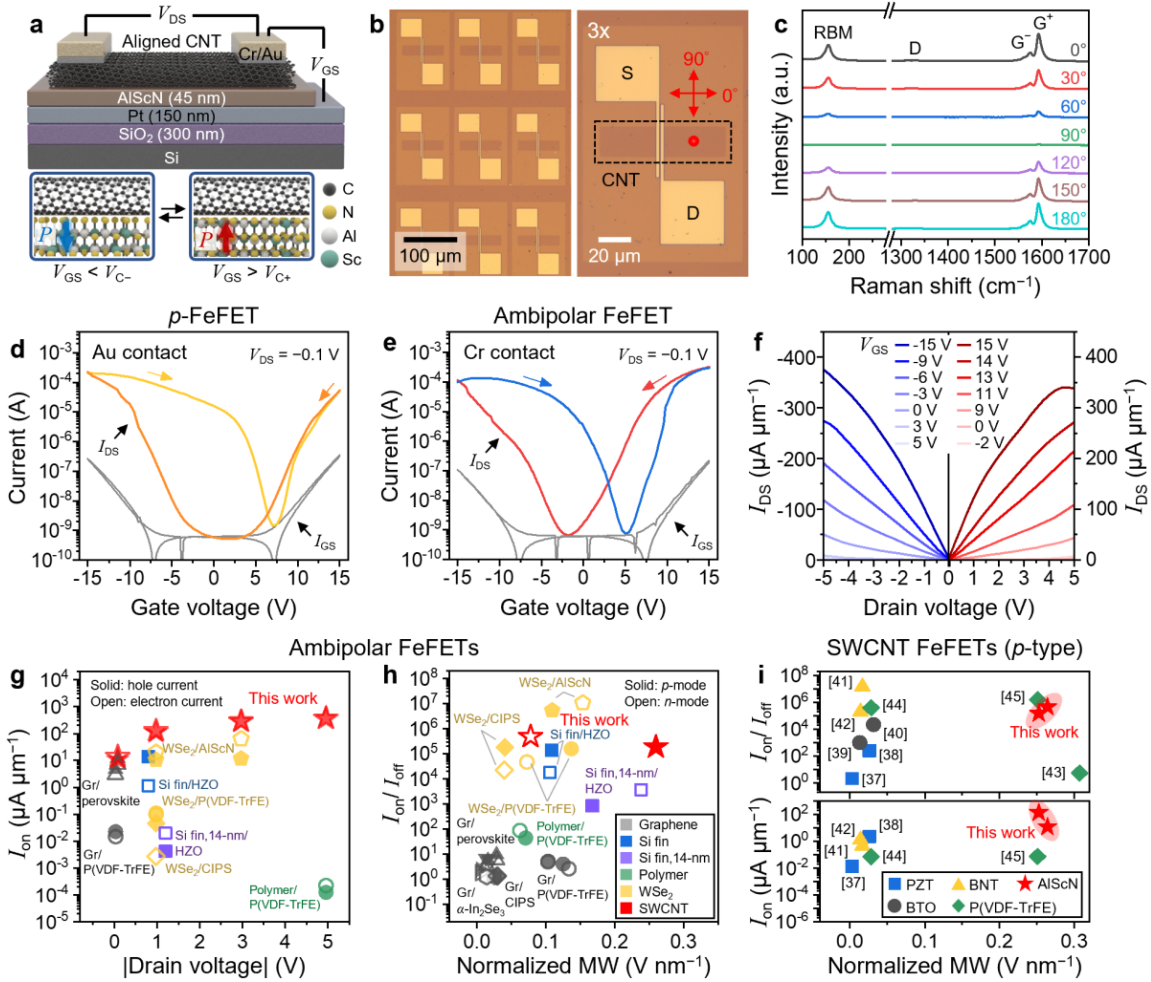

**Figure 1. Scalable FeFET arrays fabricated by integrating aligned semiconducting SWCNT channels with ferroelectric AlScN gate dielectrics.** (**a**) Scheme describing the device architecture and operating principles. (**b**) Optical microscope images of SWCNT FeFET arrays. (**c**) Angle-resolved polarized Raman spectra of the aligned SWCNT channel measured with a 633-nm laser source. Polarization angles are defined with respect to the channel length direction. Transfer characteristics of SWCNT FeFETs fabricated using (**d**) Au and (**e**) Cr contacts. (**f**) Output characteristics of ambipolar SWCNT FeFETs (Cr-contacted) for the *p*- and *n*-type operations. (**g, h**) Plots comparing the performance of ambipolar SWCNT FeFETs (this work) with previous works using various channel and ferroelectric materials: (g) on-current densities (solid data: hole current, open data: electron current) and (h) current on/off ratios and normalized memory windows (solid data: *p*-FeFET operation mode, open data: solid data: *n*-FeFET operation mode). (**i**) Plot showing the performance comparison of *p*-type SWCNT FeFETs (this work, at $V_{DS}$ of 0.1 and 1 V) with previously reported SWCNT FeFETs. Gr: graphene, HZO: $Hf_xZr_{1-x}O_2$, P(VDF-TrFE): poly(vinylidenefluoride-trifluoroethylene), CIPS: $CuInP_2S_6$, PZT: $Pb(Zr_xTi_{1-x})O_3$, BNT: $(Bi,Nd)_4Ti_3O_{12}$, BTO: $BaTiO_3$.

switches the polarization in the opposite direction and thus induces holes in the channel.

The wafer-scale, densely aligned SWCNT film with a purity close to industry-standard (99.9999%) enables scalable fabrication of FeFETs, as demonstrated by an optical microscope image of a centimeter-scale SWCNT FeFET array with the channel length ($L_{ch}$) and width ($W_{ch}$) of $L_{ch}$ = 500 nm and $W_{ch}$ = 20 μm, respectively (**Figure 1b**). Importantly, our aligned SWCNT channel retains high quality with a low level of atomic-scale defects despite the wet transfer process involved in device fabrication, which is confirmed by the intensity ratio of the Raman D peak (at 1330 cm$^{-1}$) to the G$^+$ peak (at 1594 cm$^{-1}$) of only 0.02[29]. Highly aligned nature of SWCNTs along the channel was further validated by angle-resolved polarized Raman spectra using a linearly polarized 633-nm laser source (**Figure 1c**). The intensities of Raman peaks, including radial breathing mode (RBM, 156 cm$^{-1}$) and the G$^-$ and G$^+$ bands (1572 and 1594 cm$^{-1}$), significantly decreased from their maximum to minimum as the polarization direction changed from parallel (polarization angle: 0°) to perpendicular (polarization angle: 90°) to the channel. The ratio of the G$^+$ peak intensity at 0° to that at 90° was only 0.026, indicating the highly aligned assembly of SWCNTs parallel to the channel direction.

Engineering the Fermi level of the contact metal used for source/drain electrodes enables the design of the polarity of our SWCNT FeFETs, as demonstrated by the representative transfer curves obtained by measuring the drain current ($I_{DS}$) as a function of $V_{GS}$ at a fixed drain voltage ($V_{DS}$) of −0.1 V (**Figure 1d,e**). In particular, we realized p-type FeFETs by employing metals with high work functions ($\Phi_M$) such as Au ($\Phi_M$ = 5.1 eV[54]) and Pd ($\Phi_M$ = 5.2 eV[55]), as their Fermi levels ($E_F$) before contact ($E_F$ = −5.1 eV for Au and −5.2 eV for Pd with respect to the vacuum level) are located below the valence band minimum ($E_V$) of the SWCNT channel ($E_V$ = −4.75 eV[56]) (**Figure S4**). For gate voltage sweep ranges of $V_{GS} \geq \pm 10$ V, both Au- and Pd-contacted devices exhibited clear clockwise hysteresis behavior in the hole conduction region (**Figure 1d** and **Figure S5a,b**), indicating that the ferroelectric polarization switching in the AlScN layer dominated the channel conductance[51]. The memory window (MW) and the current on/off ratio ($I_{on}/I_{off}$) reached their maximum at the largest tested sweep range of $V_{GS} = \pm 15$ V, as this sweep voltage exceeded the coercive voltage of the 45-nm AlScN layer (~13.5 V[35]) and thus allowed the ferroelectric polarization to switch between fully up and fully down states. The MW and $I_{on}/I_{off}$ ratio were 11.9 V and 4.4 × 10$^5$ for the Au-contacted devices, and 10.4 V and 2.6 × 10$^5$ for the Pd-contacted devices, with the on-current levels being similar for both types of devices. The comparison between MW and $I_{on}/I_{off}$ ratio indicates that the Au contact enhances the performance of SWCNT/AlScN devices as *p*-type FeFETs relative to the Pd contact. The *p*-type behavior and ferroelectricity-induced hysteresis were also robust at higher drain voltage for both Au and Pd contacts (**Figure S6a,b**).

In addition to *p*-type FeFETs, ambipolar FeFETs can also be fabricated using contact metals with similar Schottky barrier heights for hole and electron injection (**Figure 1d**). This approach benefits from the high remnant polarization of AlScN, which can induce substantial doping levels for both types of carriers[35,36]. To this end, we tested Ti ($E_F$ = −4.3 eV[55]) and Cr ($E_F$ = −4.5 eV[55]) due to their work functions being close to that of SWCNTs ($E_F$ = −4.5 eV[56]) and their good adhesion with low-dimensional material channels (**Figure S4**)[57-61]. Although the Ti-contacted SWCNT FeFET exhibited ambipolar transfer characteristics, hysteresis corresponding to ferroelectric switching was observed only in the electron conduction region (**Figure S5c** and **S6c**). In addition, $I_{on}$ and $I_{on}/I_{off}$ ratio were more than 2 orders of magnitude lower than those of Au- and Pd-contacted counterparts. The unexpected hysteresis behavior, along with the low $I_{on}$ and $I_{on}/I_{off}$ ratio, most likely resulted from the defective channel/electrode interface caused by a reaction between the SWCNT and Ti[62-64]. Notably, unlike the Ti-contacted device, the Cr-contacted FeFET demonstrated ambipolar transfer characteristics solely governed by ferroelectric switching (clockwise hysteresis in the hole conduction region and counterclockwise hysteresis in the electron conduction region), with the current levels similar for both hole and electron conduction. Furthermore, $I_{on}$ and $I_{on}/I_{off}$ ratio were comparable to those observed for the case of Au and Pd contacts (**Figure 1d** and **Figure S5d** and **S6d**), thus motivating the use of Cr over Ti for realizing ambipolar SWCNT FeFETs. The MW and $I_{on}/I_{off}$ ratio were 11.0 V and 1.9 × 10$^5$ in the hole conduction region, and 3.5 V and 4.8 × 10$^5$ in the electron conduction region, for the widest $V_{GS}$ sweep range tested ($V_{GS}$ = ±15 V). The ferroelectricity-dominated ambipolar characteristics were also evident at higher drain voltage (**Figure S6d**).

To further validate that the hysteresis of the SWCNT FeFETs originates from the ferroelectric switching of AlScN, we compared the current–voltage characteristics with those measured from back-gate SWCNT FETs fabricated using 50-nm SiO$_2$/Si wafers (**Figure S7**). For this comparative study, we focused on Au and Cr as representative contact metals because they achieved the best performance for *p*-type and ambipolar FeFETs, respectively. The SWCNT FeFET and FET employed identical electrode designs and channel dimensions ($L_{ch}$ = 500 nm and $W_{ch}$ = 20 μm) to directly examine the effect of 45-nm AlScN and 50-nm SiO$_2$ gate dielectrics on device characteristics. Under the gate voltage sweep of $V_{GS} = \pm 15$ V, both the Au- and Cr-contacted FETs exhibited a *p*-type behavior with charge-trapping-induced hysteresis (counterclockwise) due to the absence of ferroelectricity in the $SiO_2$ layer[35]. Thus, the comparison of hysteresis behavior between the SWCNT FeFET and FET further confirms that the electrical characteristics of the SWCNT FeFET are governed by the ferroelectric switching in the AlScN layer.

One of the major advantages of semiconducting SWCNTs for electronic applications is their ability to achieve nearly equal current levels for both electron and hole conduction, a characteristic critical for ambipolar devices but often challenging to achieve with other semiconducting materials[26-28]. By measuring the output characteristics of the ambipolar FeFETs for both *p*- and *n*-type operations, we evaluated how closely hole and electron current levels align in our SWCNT channel integrated with the AlScN layer (**Figure 1f**). Notably, the on-current densities for the *p*-type and *n*-type operation modes were similar for the tested drain voltage range (0 V to ±5 V). The on-current densities were ~100 µA µm$^{-1}$ and ~120 µA µm$^{-1}$ at $|V_{DS}| = 1$ V for hole and electron conduction, respectively, under $V_{GS}$ of −15 V and 15 V. The on-current densities increased to ~270 µA µm$^{-1}$ (hole) and ~270 µA µm$^{-1}$ (electron) at $|V_{DS}| = 3$ V, and further to ~375 µA µm$^{-1}$ (hole) and ~340 µA µm$^{-1}$ (electron) at $|V_{DS}| = 5$ V. This balanced electron and hole current levels are particularly noteworthy because silicon transistors dominating the current semiconductor industry exhibit imbalanced current densities between *p*-type and *n*-type transistors, thus necessitating additional engineering efforts such as channel width adjustments.

To ensure a fair assessment of our work, we compare the performance metrics of our SWCNT device with other reported FeFETs (**Figure 1g–i**). As illustrated in **Figure 1g**, we first compare the on-current densities, where our ambipolar SWCNT FeFET outperforms previous ambipolar FeFETs[36,65-77], including state-of-the-art silicon technologies, such as ferroelectric FinFETs at the 14-nm technology node using $Hf_xZr_{1-x}O_2$[65], and graphene-based FeFETs[68,69,71-76]. Notably, the hole (solid red star) and electron (open red star) currents of the SWCNT FeFET are well-balanced, and the on-current densities exceed those of all reported ambipolar FeFETs, reaching more than 4 times the densities achieved by the state-of-the-art ambipolar FeFETs fabricated with $WSe_2$ and AlScN[36]. In addition to the current densities, other our work demonstrates among the highest FeFET performance metrics reported for ambipolar devices, considering both the normalized MW (MW normalized by the ferroelectric thickness) and the $I_{on}/I_{off}$ ratio when operated as *p*-FeFET (solid red star) and *n*-FeFET (open red star) modes (**Figure 1h**). We also compared our device performance with FeFETs based on SWCNTs and various ferroelectric materials[37-45], such as metal oxide perovskites and poly(vinylidenefluoride-trifluoroethylene) (P(VDF-TrFE)) (**Figure 1i**). We focused our comparison on *p*-type FeFETs, as all previous SWCNT FeFETs exhibit *p*-type behavior due to the predominance of *p*-type characteristics in SWCNTs when used in transistors. Our device achieved the highest performance in terms of normalized MW and on-current density, along with the third-highest $I_{on}/I_{off}$ ratio among the reported SWCNT FeFETs. The performance comparisons clearly demonstrate the high potential of the proposed SWCNT/AlScN system for scalable, circuit-level fabrication of ambipolar FeFETs.

Benefiting from the scalability of SWCNT and AlScN films, the device characteristics of our FeFETs demonstrated excellent spatial uniformity across the centimeter-scale array (**Figure 2a–c** and **Figure S8–S9**). As shown in **Figure 2a**, transfer curves measured from different ambipolar FeFETs (Cr contact) within the array closely match and overlap with each other. The *p*-type FeFETs (Au contact) also exhibited a good uniformity in transfer characteristics across the array (**Figure S8**). The device-to-device uniformity of ambipolar FeFETs was further investigated by performing statistical analyses of the key performance metrics, including MW, voltages corresponding to the minimum current in the high-resistance states ($V_{HRS}$) in both the hole and electron conduction regions, and the current ratio of low-resistance state (LRS) to high-resistance state (HRS) measured at $V_{HRS}$ ($I_{LRS}/I_{HRS}$). The average and standard deviation (SD) of MW were 11.8 V and 0.9 V for hole conduction side and 3.8 V and 0.6 V for electron conduction side (**Figure S9**). $V_{HRS}$ were also narrowly distributed, with average values of −1.8 V (SD: 0.5 V) for hole conduction side and 5.9 V (SD: 0.5 V) for electron conduction side (**Figure 2b**). The average $I_{LRS}/I_{HRS}$ values corresponding to these operation voltages were $4.7 \times 10^4$ (SD: $2.3 \times 10^4$) and $3.0 \times 10^3$ (SD: $2.4 \times 10^3$) for hole and electron conduction sides, respectively (**Figure 2c**).

Next, we assessed the retention characteristics of the ambipolar SWCNT FeFET as a non-volatile memory device by measuring the time evolution of current levels in both LRS and HRS across hole and electron conduction regimes (**Figure 2d**). Prior to the retention measurement, each state was set by applying a gate voltage pulse of $|V_{GS}| = 15$ V with a pulse width of 1 s to achieve the desired polarization. In specific, $V_{GS} = 15$ V was used to achieve upward polarization for LRS in the electron conduction region and HRS in the hole conduction region, while $V_{GS} = −15$ V was applied to induce downward polarization for LRS in the hole conduction region and HRS in the electron conduction region. The drain current at different time points was then read by applying $V_{GS} = -2$ V and $V_{DS} = -0.1$ V for the memory states in the hole conduction region (**Figure 2e**), and $V_{GS} = 5$ V and $V_{DS} = 0.1$ V for the memory states in the electron conduction region (**Figure 2f**). Remarkably, the LRS and HRS in both the hole and electron conduction regions were characterized by stable retention behavior for up to $10^4$ s after the initial program/erase pulse.

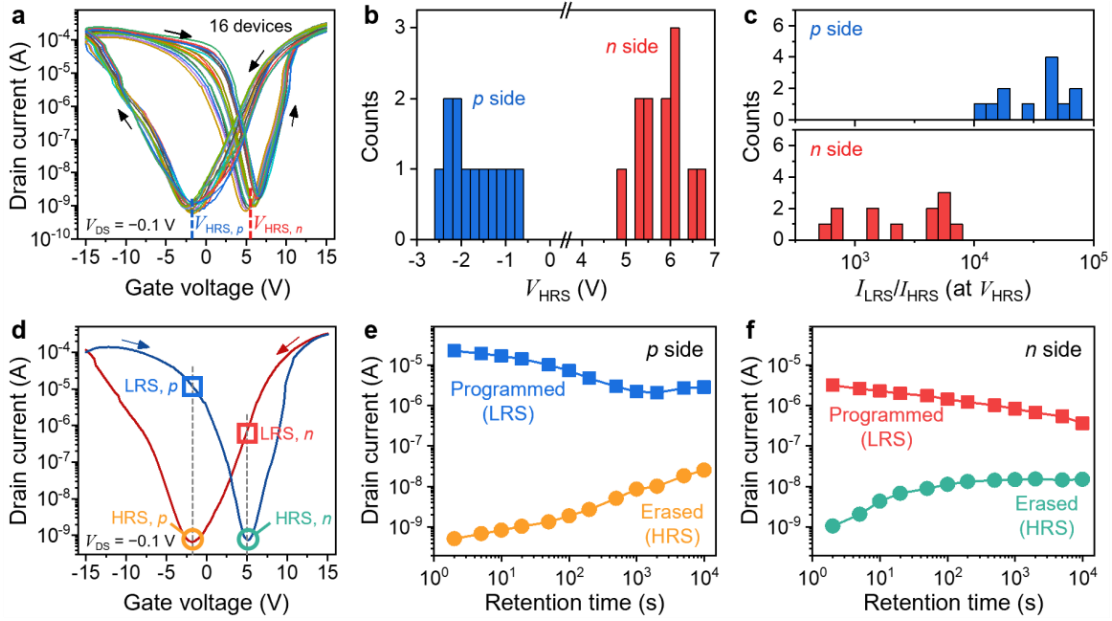

**Figure 2. Memory Characteristics of SWCNT FeFETs.** (a) Transfer curves of 16 devices in the FeFET array. Histograms showing (b) HRS voltages and (c) $I_{LRS}/I_{HRS}$ ratios at HRS voltages. (d) Transfer curves of the device used for retention measurements ($V_{GS}$ sweep range: ±15 V). Retention characteristics of memory states in (e) hole conduction and (f) electron conduction sides.

In addition to serving as memory devices, our ambipolar SWCNT array enables the realization of FETs with switchable polarities, which is greatly beneficial for reducing the number of device components needed to construct circuits compared to unipolar FETs, as well as for realizing multiple circuits through reconfiguration (**Figure 3**). By programming the polarization direction of AlScN with a gate bias ($V_{GS} = \pm 14$ V) and operating within a $V_{GS}$ range smaller than the coercive voltage, we achieved switching between p-channel (polarization opposite to the channel) and n-channel (polarization toward the channel) FETs using a single SWCNT device (**Figure 3a**). **Figure 3b,c** shows transfer curves measured over 100 cycles of device polarity reconfiguration. The stability of the SWCNT channel and AlScN layer during reconfiguration was demonstrated by the nearly consistent transfer curves. The device exhibited similar $I_{on}$ and $I_{on}/I_{off}$ ratio for p-channel and *n*-channel FET operations, highlighting the advantage of the SWCNT channel in achieving symmetry between the two modes (**Figure 3c,d**). Furthermore, the $I_{on}/I_{off}$ ratio remained above 3 × $10^3$ throughout the reconfiguration cycles, which is sufficient for constructing digital circuits ($I_{on}/I_{off} > 10^3$)[78,79].

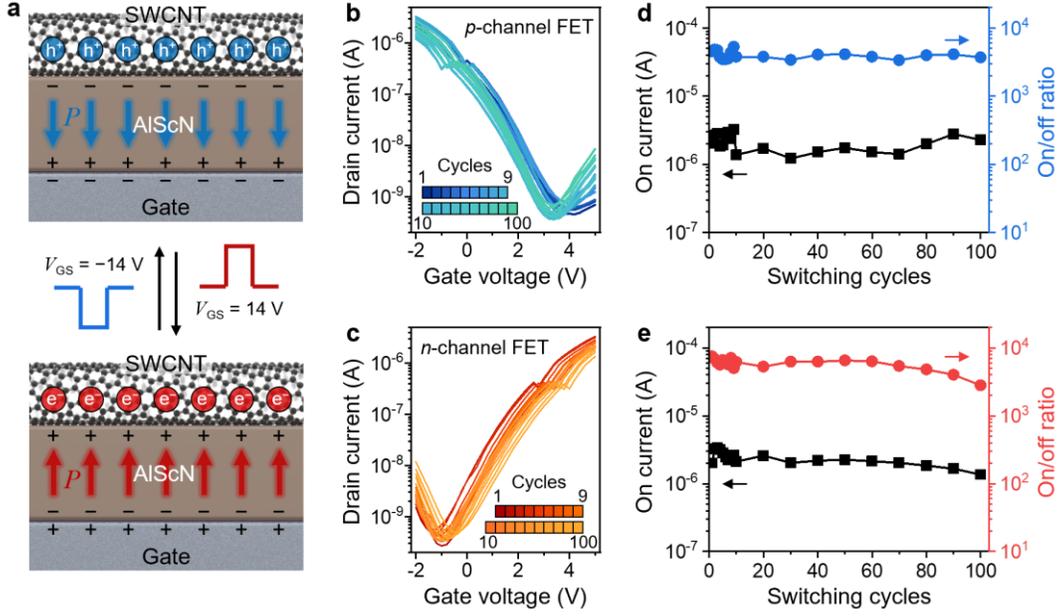

**Figure 3. Reconfigurable *p*- and *n*-type FETs realized by SWCNT FeFETs. (a)** Scheme describing the reconfiguration strategy. Transfer curves measured during **(b)** *p*-channel FET and **(c)** *n*-channel FET reconfiguration up to 100 cycles. **(d-e)** Corresponding on-currents and on/off ratios.

The experimental results presented thus far demonstrate that our SWCNT FeFET offers a variety of multifunctional capabilities, including *p*-polar and ambipolar NVM devices, as well as *p*- and *n*-type polarity switchable FETs. In this section, we show the practical utility of the SWCNT FeFET in circuit applications, specifically by demonstrating its use in ternary content-addressable memory (TCAM). Notably, as will be compared in more detail, a single ambipolar SWCNT FeFET provides the full functionality required for a non-volatile TCAM, in contrast to conventional CMOS based TCAM designs, which typically require 10 or more transistors[80].

To validate this, we performed circuit simulations using Cadence Virtuoso. The TCAM circuit comprises an SWCNT FeFET, a PMOS transistor, and a pre-charge capacitor (**Figure 4a**). The SWCNT FeFET serves as the core-cell of the TCAM (whereas conventional silicon-based TCAM requires at least 10 transistors as will be discussed), while the PMOS transistor charges the pre-charge capacitor, which is used to evaluate whether the input to the search line (SL) matches the stored state in the core SWCNT FeFET. It is important to note that the PMOS transistor and pre-charge capacitor are peripheral components to the TCAM core-cell, and multiple TCAM core-cells are typically connected to the peripheral circuitry.

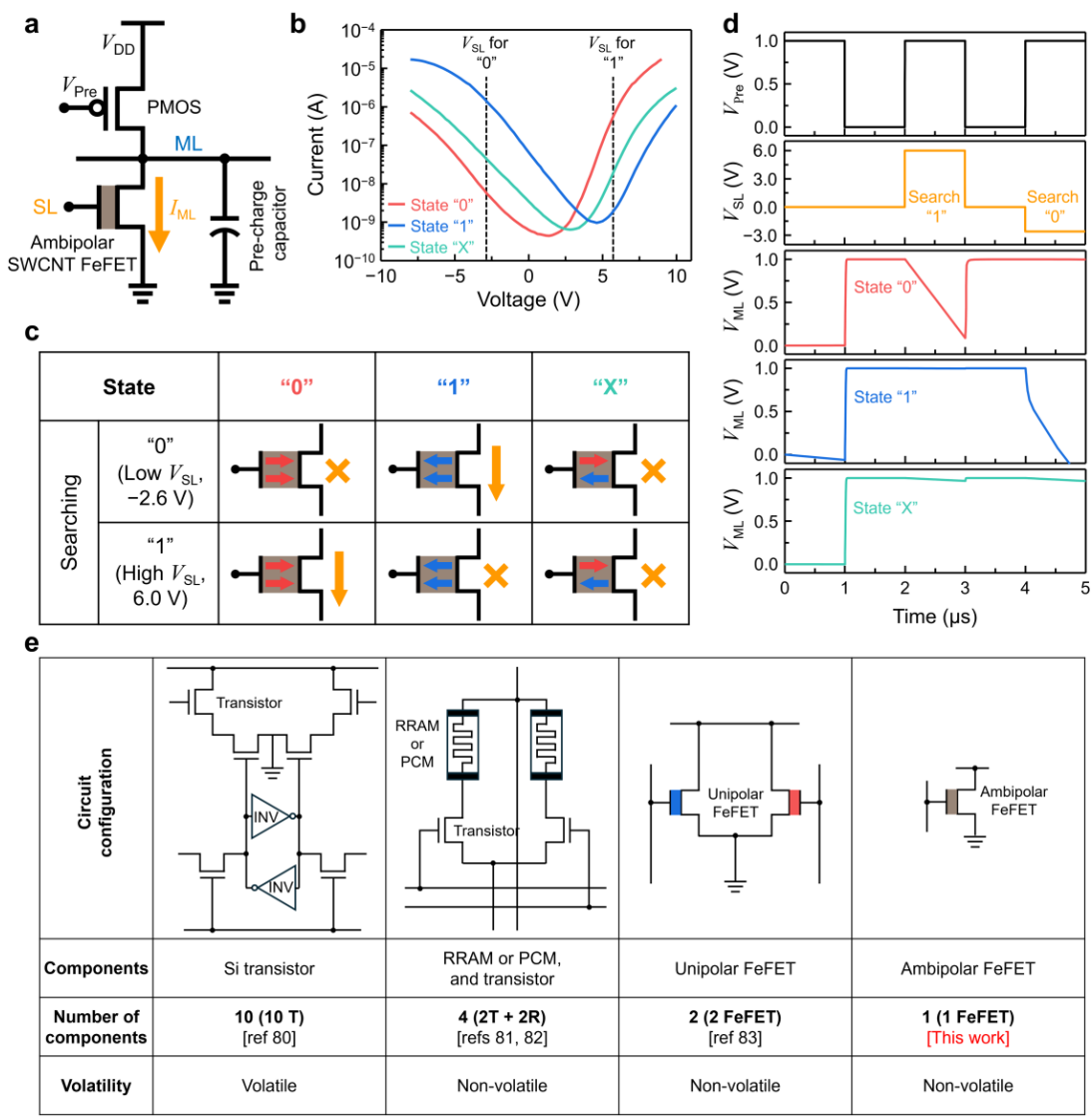

**Figure 4. Ternary content-addressable memory (TCAM) using single SWCNT FeFET.** (**a**) Schematic illustration of the TCAM circuit based on an ambipolar SWCNT FeFET. (**b**) Transfer curves of the ambipolar FeFET, showing the distinct behavior in different polarization states: "0" (red), "1" (blue), and "don't care" ("X") (green) states. (**c**) Scheme describing the channel conduction across all possible cases, demonstrating how the match or mismatch between the search voltage ($V_{SL}$) and the stored state in the FeFET affects conduction. (**d**) Simulation results reflecting the behavior of the TCAM for each matching and mismatching scenario, with corresponding voltage outputs for the match line (ML). (**e**) A comparison table showcasing various TCAM circuit configurations, including conventional silicon-based transistors, RRAM or PCM-based systems, and FeFET-based designs. The table highlights the significant reduction in component count achieved with the ambipolar FeFET, which requires only a single device for full TCAM functionality, thus representing the most compact non-volatile solution.

**Figure 4b** presents the experimental transfer curves of the ambipolar SWCNT FeFET, where each curve exhibits a different threshold voltage due to shifts caused by varying degrees of ferroelectric polarization. We define the curve with the leftmost threshold voltage (red) as state "0," the one with the rightmost threshold voltage (blue) as state "1," and the middle curve (green) as the "don't care" state, "X." Note that the leftmost and rightmost threshold voltages differ slightly from those in **Figure 3b,c**, as a wider gate voltage range ($V_G$ from −8 to 10 V, instead of −2 to 5 V) was swept to achieve sufficiently high contrast between current levels for the three states. Sweeping this wider voltage range might have induced partial polarization of the pre-polarized AlScN, resulting in shifts in threshold voltages compared to when a narrower gate voltage sweep range was used. For reliable circuit simulations, we emphasize that the experimentally obtained transfer curves were directly used to create an accurate model within Cadence. As shown in **Figure S10**, this model precisely replicates the experimental transfer curves and serves as the basis for the TCAM demonstration.

To provide a comprehensive understanding of the TCAM operation, we schematically illustrate different scenarios in **Figure 4c** based on the input and stored states of the ambipolar FeFET. When the stored state and the search voltage ($V_{SL}$) match—meaning the FeFET state is "0" (or "1") and $V_{SL}$ is likewise "0" (or "1")—the current through the FeFET channel is low. Conversely, when the stored state of the FeFET and $V_{SL}$ do not match, the current is high. In cases where the "don't care" state ("X") is stored, the current remains low regardless of whether $V_{SL}$ is "0" and "1," enabling full TCAM functionality. These current levels, as functions of the FeFET state and $V_{SL}$, are visually illustrated in **Figure 4b** for better clarity.

Based on this operational principle, a full TCAM circuit simulation is performed as shown in **Figure 4d** and **Figure S11**. The topmost curve (black) in **Figure 4d** represents the pre-charge voltage ($V_{Pre}$), which is applied to charge the pre-charge capacitor to 1 V. The orange curve shows the applied $V_{SL}$, while the red, blue, and green curves represent the voltage on the match line (ML) corresponding to the stored states of the FeFET: "0," "1," and "X." Depending on whether the stored state matches or mismatches with the applied $V_{SL}$, the voltage across the pre-charged capacitor is either discharged to ground or maintained. In this simulation, a 1 pF capacitor is used, yielding a discharge time ($t$) of 1 μs, which is determined using the equation $I \times t = C \times V$, where *I* is the current, *C* is the capacitance, and *V* is the voltage. When the stored state and $V_{SL}$ do not match, the current through the FeFET is approximately 1 µA, and with a 1 V charge across the capacitor, the discharge time is around 1 µs with a 1 pF capacitor. In contrast, when the stored state and $V_{SL}$ match, the current is two orders of magnitude lower than the mismatched case, resulting in a discharge time that is more than 100 times longer, allowing for a clear distinction between the stored "0" and "1" states. In the case of the "X" state, the current is more than an order of magnitude lower than 1 µA, which yields a discharge time more than 10 times longer compared to the mismatched case. This difference in discharge times enables reliable TCAM operation with clear state distinction, as reflected in **Figure 4d**. The charge in the capacitor is quickly discharged, lowering the ML voltage only when the stored state and $V_{SL}$ do not match. Conversely, when the stored state matches $V_{SL}$ or is "X", the capacitor charge and thus the ML voltage are retained.

This demonstration of full TCAM functionality using a single ambipolar FeFET as the core component represents a significant advancement in reducing the number of components required, especially when compared to conventional silicon CMOS-based TCAM designs. As shown in **Figure 4e**, traditional TCAM configurations typically require more than 10 transistors[80]. Recent developments using alternative technologies, such as resistive random-access memory (RRAM)[81], phase-change memory (PCM)[82], or even unipolar FeFETs[83,84], have managed to lower the number of components to around 4 or 2 components. However, the ambipolar FeFET-based TCAM presented here stands out as the most compact design, requiring only a single ambipolar FeFET (**Figure 4e**), which significantly simplifies the circuit architecture while maintaining functionality. This reduction in component count could offer substantial benefits in terms of integration density and power efficiency when scaled to large arrays and implemented in highly miniaturized devices.

## Conclusion

In summary, we have demonstrated a scalable array of single-gate reconfigurable devices capable of switching between *p*-type and *n*-type FETs as well as NVMs, based on CMOS BEOL-compatible SWCNT/AlScN FeFETs. The strong spontaneous polarization of AlScN, combined with the excellent ambipolarity of the SWCNT channel, enabled high and balanced on-state current densities (~270 µA µm$^{-1}$ at a drain voltage of 3 V) and wide memory windows (0.26 V nm$^{-1}$ and 0.08 V nm$^{-1}$, respectively) in both the hole and electron conduction regimes. These performance metrics are among the highest reported for ambipolar FeFETs with different channel materials and ferroelectric combinations. Leveraging the reliable reconfiguration between *p*-type and *n*-type FETs, a TCAM cell could be constructed with significantly fewer devices compared to conventional MOSFETs with fixed polarities and other recent technologies, highlighting the potential of SWCNT FeFETs for miniaturized, low-power circuit applications. Our work promises the realization of high-performance reconfigurable devices based on highly scalable 1D semiconductors and III-nitride ferroelectric films, creating opportunities for post-Moore's Law innovations.

## Methods

### Production of aligned semiconducting SWCNT films

Aligned semiconducting SWCNT films were produced based on a procedure reported in a previous work[29]. First, high-purity (>99.9999%) semiconducting SWCNT dispersions were obtained by sorting arc-discharged CNT powders, which consist of a mixture of semiconducting and metallic CNTs, as well as metallic catalysts used for CNT synthesis. Commercially purchased arc-discharged CNT powders (AP-SWNT from Carbon Solutions, Inc.) were dissolved in in toluene with the presence of conjugated poly[9-(1-octylonoyl)-9H-carbazole-2,7-diyl] (PCz) molecules as dispersants, followed by multiple iterations of ultracentrifugation, rinsing with tetrahydrofuran, and redispersion in 1,1,2-trichloroethane to enrich semiconducting SWCNTs with diameters narrowly distributed around 1.5 nm. Next, the resulting SWCNTs were assembled into an aligned film using dip coating in the presence of interfacial confinement. Specifically, after vertically dipping a 4-inch 300 nm $SiO_2$/Si wafer into the dispersion, 2-butene-1,4-diol, which is immiscible with the subphase 1,1,2-trichloroethane but can interact with the PCz-wrapped SWCNTs, was spread around the interface between the wafer, 1,1,2-trichloroethane, and air. Because SWCNTs in the 1,1,2-trichloroethane did not stick to the wafer surface, only the SWCNTs confined at the 1,1,2-trichloroethane/2-butene-1,4-diol/wafer interface were deposited on the wafer during the withdrawal stage while SWCNTs were continually supplied from the 1,1,2-trichloroethane subphase. The wafer was withdrawn at a speed of 2 μm s$^{-1}$, resulting in a dense, monolayer film of highly aligned SWCNTs.

### AlScN growth

AlScN ($Al_{0.68}Sc_{0.32}N$) films were deposited on 4-inch Si (100) wafers coated with 300 nm of $SiO_2$,

a 10 nm Ti seed/adhesion layer, and a 150 nm Pt film with a (111) crystal orientation (Pt/Ti/SiO$_2$/Si) (MTI Corporation) using a pulsed DC deposition system (CLUSTERLINE® 200 II, Evatec). In specific, the AlScN deposition processes were performed via 150 kHz pulsed DC co-sputtering of Al and Sc targets with 20 sccm N$_2$ flow under $8.3 \times 10^{-4}$ mbar in the sputtering chamber. The chamber temperature was maintained at 350 °C, which is a back-end-of-line (BEOL)-compatible thermal budget. The AlScN/Pt/Ti/SiO$_2$/Si substrates used in this work were from the same wafer used in our previous works[35,36].

**Devices fabrication**

To fabricate ferroelectric field-effect transistors, the aligned SWCNT film was first transferred from the 300 nm SiO$_2$/Si substrate onto an AlScN/Pt/Ti/SiO$_2$/Si substrate using the poly(methyl methacrylate) (PMMA)-assisted wet transfer method[52,53]. Specifically, a 640-nm thick PMMA support layer was coated on a ~1 cm$^2$ piece of the aligned SWCNT/300 nm SiO$_2$/Si wafer by spin coating a 8% PMMA solution in anisole (495 PMMA A8, Kayaku Advanced Materials Inc.) at 2500 rpm for 60 s, followed by baking on a hot plate at 180 °C for 5 minutes. The PMMA/SWCNT bilayer film was then released from the substrate by etching the SiO$_2$ layer with 2M KOH (Sigma Aldrich) at 60 °C for approximately 90 minutes. The PMMA/SWCNT stack was floated sequentially on three baths of deionized water to rinse off residual KOH from the surface of SWCNT and then transferred on a 2:1 volume mixture of water and isopropanol (IPA). The addition of IPA was critical for reducing the surface tension of the liquid, enabling the transfer of PMMA/SWCNT onto the hydrophobic surface of AlScN[85,86]. After etching the aluminum capping layer from the AlScN/Pt/Ti/SiO$_2$/Si substrate using 1% hydrofluoric acid, the PMMA/SWCNT film was immediately transferred on the AlScN surface by dipping the substrate into the water/IPA mixture and scooping the bilayer stack. Following solvent evaporation, the PMMA layer was removed from the SWCNT film using acetone and IPA. Subsequently, the sample was annealed at 350 °C with a 100 sccm H$_2$/Ar (15 % H$_2$, balance Ar) flow for 2 hours to further remove residual PMMA.

Next, the electrode and channel fabrication was carried out using electron-beam lithography with a bilayer PMMA resist. The resist coating process involved spin coating a 4% PMMA solution in anisole (495 PMMA A4, Kayaku Advanced Materials Inc.) at 2500 rpm for 60 seconds and baking at 180 °C for 5 minutes, followed by spin coating an 8% PMMA solution (495 PMMA

A8, Kayaku Advanced Materials Inc.) at 2500 rpm for 60 seconds and a second baking at 180 °C for 5 minutes. To define source and drain electrode regions, the PMMA layer was patterned using an electron beam accelerated at 50 kV in an electron-beam lithography system (Elionix ELS-7500EX, Elionix Inc.) and developed in a 1:3 volume ratio mixture of methyl isobutyl ketone (MIBK)/IPA for 40 seconds. Then, contact metals were deposited using an electron-beam evaporation tool (PVD 75 electron-beam evaporator, Kurt J. Lesker Company). For gold and palladium contact devices, 50 nm of gold or palladium was deposited at a rate of 0.2 Å s$^{-1}$ for the first 20 nm and then at a rate of 2 Å s$^{-1}$ for the remaining 30 nm. Chromium and titanium contact devices were fabricated by depositing 20 nm of chromium or titanium at a rate of 0.2 Å s$^{-1}$, followed by the deposition of a 30 nm gold capping layer at a rate of 2 Å s$^{-1}$ to prevent oxidation of the contact metal upon exposure to ambient air. After the metal deposition, the PMMA layer was removed from the SWCNT film via lift-off using acetone, resulting in an array of source and drain electrodes in areas that were not covered by the patterned PMMA. To define channels in the SWCNT film, another bilayer PMMA resist was spin-coated onto the sample using 4% and 8% PMMA solutions, followed by electron-beam lithography and development in a 1:3 mixture of MIBK/IPA. SWCNTs in regions outside the PMMA patterns were etched by performing reactive ion etching (Jupiter II tabletop reactive ion etcher, March Instruments) with oxygen gas (flow rate: 90 sccm) at a power of 100 W for 40 s. Then, PMMA was removed from the surface using acetone to complete the device fabrication.

**Electrical characterization of devices**

Electrical measurements were conducted at room temperature using a probe station connected to a semiconductor parameter analyzer (Keithley 4200A-SCS parameter analyzer, Tektronix Inc.). Current-voltage characteristics of devices in ambient conditions were analyzed using a Cascade Microtech MPS 150 probe station (FormFactor Inc.), while those under vacuum were measured using an HP1000V-MPS+ probe station (Instec Inc.).

**Raman spectroscopy**

Polarized Raman spectra of aligned SWCNT films were obtained using a confocal Raman spectrometer (LabRAM HR Evolution, Horiba Scientific) equipped with a 633 nm laser excitation source and a 100× objective lens. The incident laser from the source was linearly polarized using a polarizer, with the polarization angle controlled by a half-wave plate. Samples were excited with a power below 2 mW to minimize damage to SWCNTs. Raman scattering signals from the sample were dispersed by a diffraction grating with 600 grooves per millimeter and recorded by a charged-coupled device (CCD) detector.

## Microscopy

Optical microscope images of devices were obtained using a bright-field optical microscope (Olympus BX51, Olympus Corporation). The surface topography of aligned SWCNT films was characterized using an atomic force microscope (Bruker ICON, Bruker Corporation) operated in tapping mode.

## Acknowledgements

This work was primarily supported by the Air Force Office of Scientific Research GHz-THz program (FA9550-23-1-0391). D.R. acknowledges primary support from the Korea Institute for Advancement of Technology (KIAT) grant funded by the Korea Government (MOTIE) (P0017305, Human Resource Development Program for Industrial Innovation (Global)) and BrainLink program funded by the Ministry of Science and ICT through the National Research Foundation of Korea (RS-2023-00237308). S.S. acknowledges support from the Institute for Basic Science (IBS-R036-D1). This work was carried out in part at the Singh Center for Nanotechnology at the University of Pennsylvania, which is supported by the NSF National Nanotechnology Coordinated Infrastructure Program under grant NNCI-2025608.

## Author contributions

D.J. supervised all aspects of the project and collaboration. D.R., K.-H.K., D.J., and R.H.O. conceived the idea of integrating large-area SWCNTs with AlScN for the scalable fabrication of unipolar and ambipolar FeFET arrays, as well as for the realization of reconfigurable device applications. D.R. and K.-H.K. designed the experiments and performed device fabrication. D.R. conducted materials characterization of SWCNT channels, including AFM and polarized Raman spectroscopy. K.-H.K. performed circuit simulations for the application of reconfigurable SWCNT FeFETs in TCAM. D.R., K.-H.K., and S.S. conducted current-voltage measurements, and retention tests. R.H.O. supervised the growth process of AlScN films and J.Z. performed AlScN deposition and characterization. L.-M.P. provided films of aligned semiconducting SWCNTs. D.R., K.-H.K., and D.J. analyzed the data, prepared figures, and wrote the manuscript. J.K. supervised materials integration and device characterization. All the authors contributed to the discussion/analysis of the results and manuscript writing.

## Competing interests

The authors declare no competing interests.

Supplementary Information

# Reconfigurable SWCNT ferroelectric field-effect transistor arrays

Dongjoon Rhee,[1,2,7] Kwan-Ho Kim,[1,7] Jeffrey Zheng,[3] Seunguk Song,[1,4,5] Lian-Mao Peng,[6] Roy H. Olsson III,[1] Joohoon Kang,[2,*] and Deep Jariwala[1,*]

[1]Department of Electrical and Systems Engineering, University of Pennsylvania, Philadelphia, PA, USA

[2]School of Advanced Materials Science and Engineering, Sungkyunkwan University (SKKU), Suwon 16419, Republic of Korea

[3]Department of Materials Science and Engineering, University of Pennsylvania, Philadelphia, PA, USA

[4]Department of Energy Science, Sungkyunkwan University (SKKU), Suwon 16419, Republic of Korea

[5]Center for 2D Quantum Heterostructures, Institute for Basic Science (IBS), Sungkyunkwan University (SKKU), Suwon 16419, Republic of Korea

[6]Center for Carbon-Based Electronics, School of Electronics, Peking University, Beijing, China

[7]These authors contributed equally: Dongjoon Rhee, Kwan-Ho Kim

*Corresponding author: Deep Jariwala (email: dmj@seas.upenn.edu), Joohoon Kang (email: joohoon@skku.edu)

**Device fabrication process**

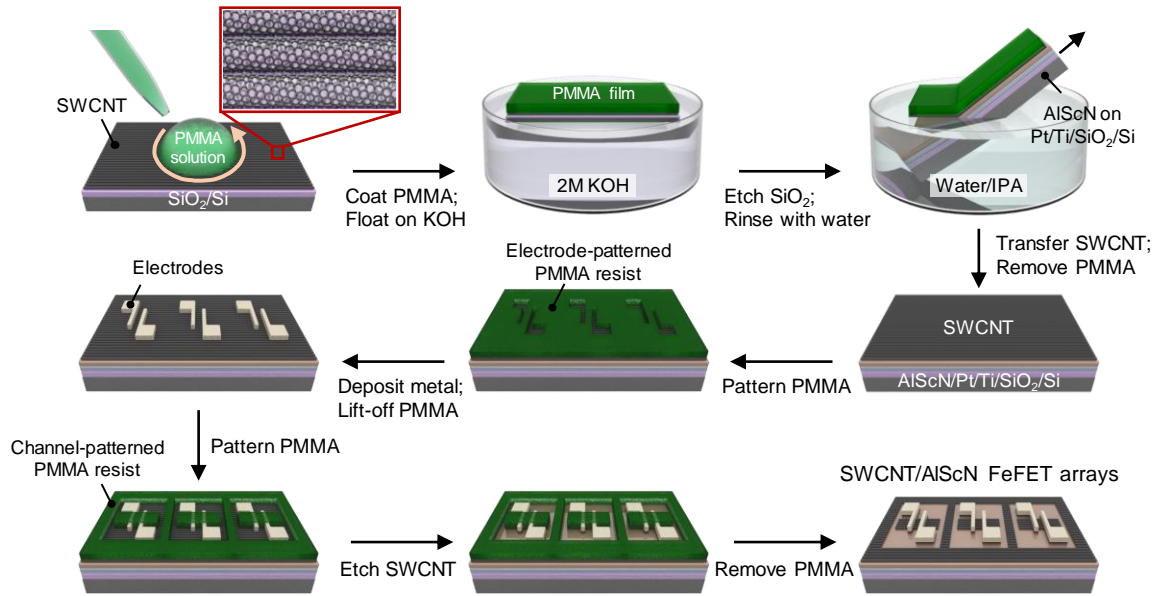

**Figure S1. Overall process to fabricate SWCNT FeFET arrays.**

**Surface topography of aligned SWCNT film**

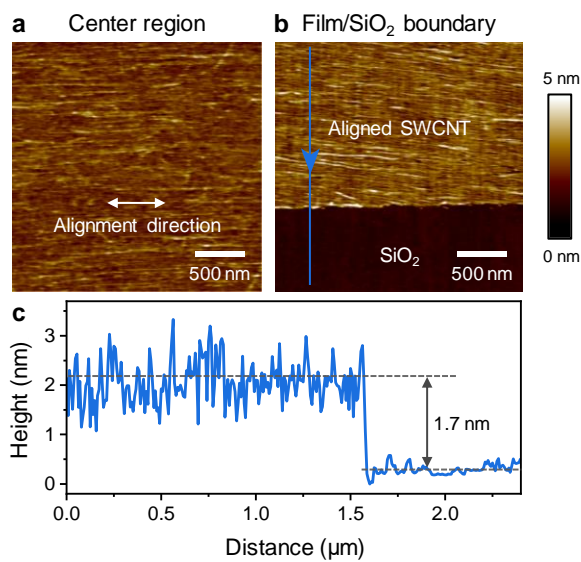

**Figure S2. Aligned SWCNT film assembled on a SiO$_2$/Si wafer before transfer.** AFM topography images of the film **(a)** in the center region and **(b)** at the boundary between the film and SiO$_2$. **(c)** Height profile along the blue line marked in the AFM image in panel (b). The measured film thickness (~1.7 nm) confirms the monolayer assembly of SWCNTs, considering that the average diameter of the SWCNTs used in this study was ~1.5 nm and that AFM tends to overestimate the thickness.

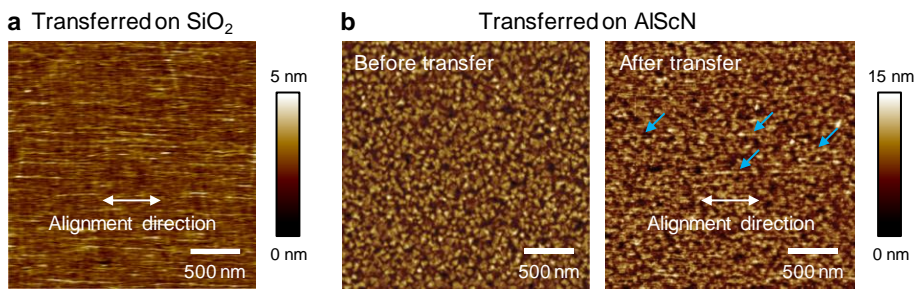

**Figure S3. Aligned SWCNT film after transfer.** AFM topography images of the film transferred on **(a)** 50-nm SiO$_2$ film thermally grown on a Si wafer and **(b)** 45-nm AlScN film deposited on a Pt/Ti/SiO$_2$/Si wafer.

**Expected energy levels for SWCNT channel and contact metals**

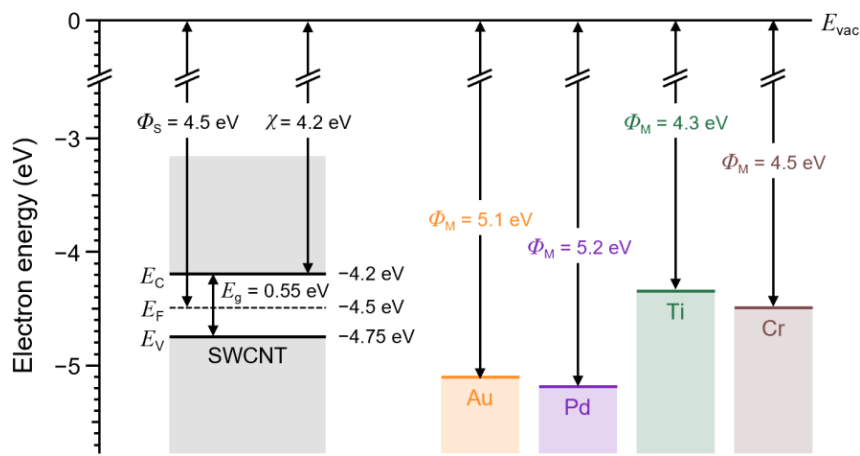

**Figure S4**. **Band diagram depicting expected energy level alignment of SWCNT channel and contact metals before contact.**

**Transfer curves of SWCNT FeFETs and FETs measured in ambient**

**Supplementary Note 1. Transfer characteristics under different $V_{GS}$ sweep ranges**

The use of high work function contact metals with a Fermi level lower than the valence band minimum of the SWCNT channel ($E_V = -4.75$ eV[1] with respect to the vacuum level), such as Au ($E_F = -5.1$ eV[2]) and Pd ($E_F = -5.2$ eV[3]), allowed us to fabricate *p*-type FeFETs (**Figure S4**). We observed that the hysteresis behavior depended on the range of the $V_{GS}$ sweep applied to measure transfer curves. When the range of gate voltage sweep was narrow ($V_{GS} \leq \pm 5$ V at $V_{DS} = -0.1$ V), Au- and Pd-contacted SWCNT FeFETs exhibited a small, counterclockwise hysteresis in the transfer curves (**Figure S5a,b**). This counterclockwise hysteresis is attributed to charge trapping and detrapping processes associated with residual water and oxygen molecules at the interface between the SWCNT channel and AlScN layer possibly introduced during the aqueous solution-based etching of the aluminum capping layer and transfer process[4-6]. As the gate voltage sweep range widened to $V_{GS} = \pm 8$ V, the direction of hysteresis changed to clockwise due to the onset of partial ferroelectric switching in AlScN[7]. The ferroelectricity-induced hysteresis became more pronounced as the gate voltage sweep range increased ($V_{GS} \geq \pm 10$ V). Largest MW and $I_{on}/I_{off}$ ratio were achieved with a $V_{GS}$ sweep range of $\pm 15$ V, where the ferroelectric polarization could switch between fully up and fully down states since the voltage exceeded the coercive voltage of the 45-nm AlScN layer (~13.5 V[7]).

We also investigated Ti ($E_F = -4.3$ eV[3]) and Cr ($E_F = -4.5$ eV[3]) as contact metals with Fermi levels near the midpoint of the bandgap of the SWCNT channel (**Figure S4**), aiming to realize ambipolar FeFETs. Notably, both Ti- and Cr-contacted SWCNT FeFETs exhibited *p*-type transfer curves with counterclockwise hysteresis for a narrow gate voltage sweep of $V_{GS} = \pm 5$ V at $V_{DS} = -0.1$ V (**Figure S5c,d**). This *p*-type behavior, rather than ambipolar, can be attributed to charge trapping and detrapping by residual water and oxygen molecules at the SWCNT/AlScN interface, which induce hole doping and also suppress electron conduction through electrostatic screening of the positive gate field[4,8]. As the $V_{GS}$ sweep range increased further, the device polarity gradually transitioned to ambipolar, with ferroelectric switching beginning to dominate the hysteresis behavior. For Ti-contacted SWCNT FeFETs, however, ferroelectric switching was evident only in the electron conduction region, as indicated by the hysteresis direction in the transfer curves (**Figure S5c**). In addition, both $I_{on}$ and $I_{on}/I_{off}$ ratio were more than 2 orders of magnitude lower compared to SWCNT FeFETs with other metal contacts. The unexpected hysteresis direction in the hole conduction region, combined with the low $I_{on}$ and $I_{on}/I_{off}$ ratio, may have originated from the defective channel/electrode interface, possibly formed by a reaction between the SWCNT and Ti[9-11]. In contrast, Cr-contacted FeFETs exhibited ambipolar behavior described by ferroelectric switching both in the hole (clockwise hysteresis) and electron (counterclockwise hysteresis) conduction regions. (**Figure S5d**). Furthermore, the device achieved similar current levels in both regions, with $I_{on}$ and $I_{on}/I_{off}$ ratio comparable to those of Au- and Pd-contacted counterparts.

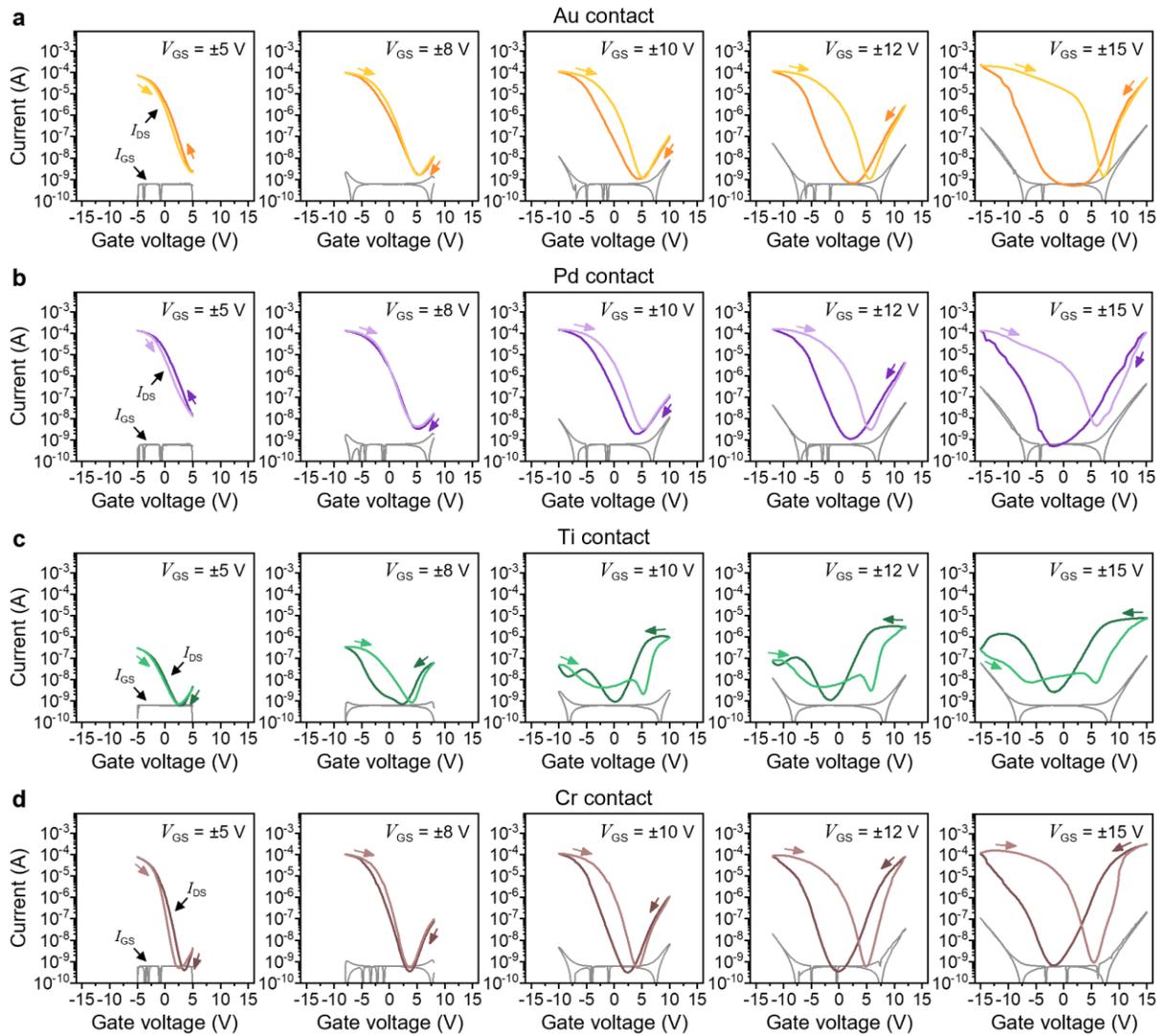

**Figure S5. Transfer curves of SWCNT FeFETs fabricated using different contact metals measured under different gate voltage sweep ranges.** (**a**) Au, (**b**) Pd, (**c**) Ti, and (**d**) Cr contacts. Drain voltage was fixed at $V_{DS} = -0.1$ V.

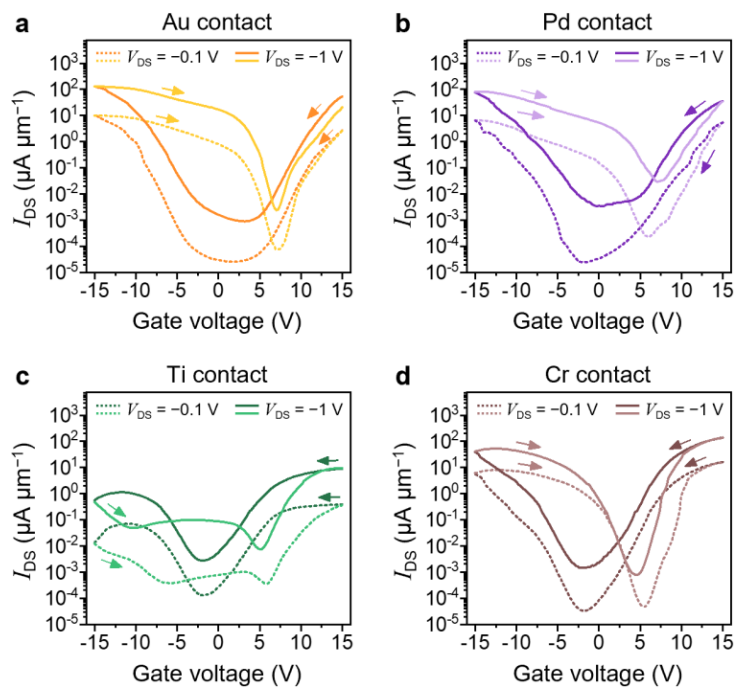

**Figure S6. Transfer curves of SWCNT FeFETs fabricated using different contact metals measured at different drain voltages. (a)** Au, **(b)** Pd, **(c)** Ti, and **(d)** Cr contacts. Gate voltage sweep range was $V_{GS} = \pm 15$ V for all measurements.

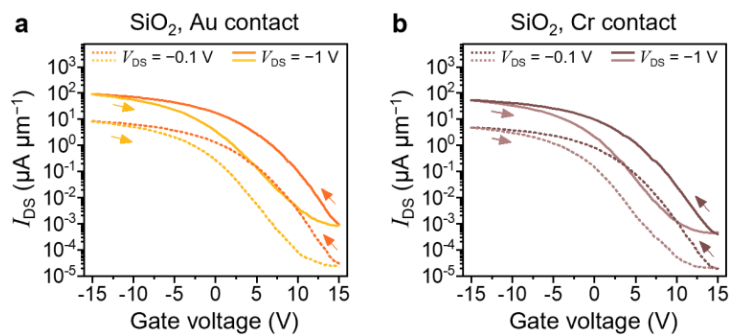

**Figure S7. Transfer curves of SWCNT FETs fabricated using 50-nm SiO$_2$ gate dielectric and different contact metals measured at different drain voltages. (a)** Au and **(b)** Cr contacts. Gate voltage sweep range was $V_{GS} = \pm 15$ V for all measurements.

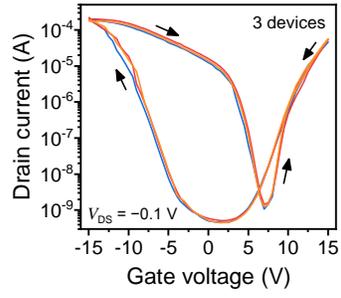

**Figure S8. Transfer curves of *p*-type SWCNT FeFETs measured over multiple devices within the array.** Au was used as the contact metal.

**Device-to-device uniformity in the memory window of SWCNT FeFETs**

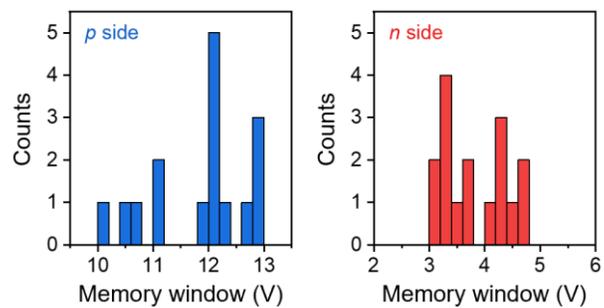

**Figure S9**. **Histogram showing memory windows of 16 devices in the array.**

**TCAM simulations using Cadence Virtuoso**

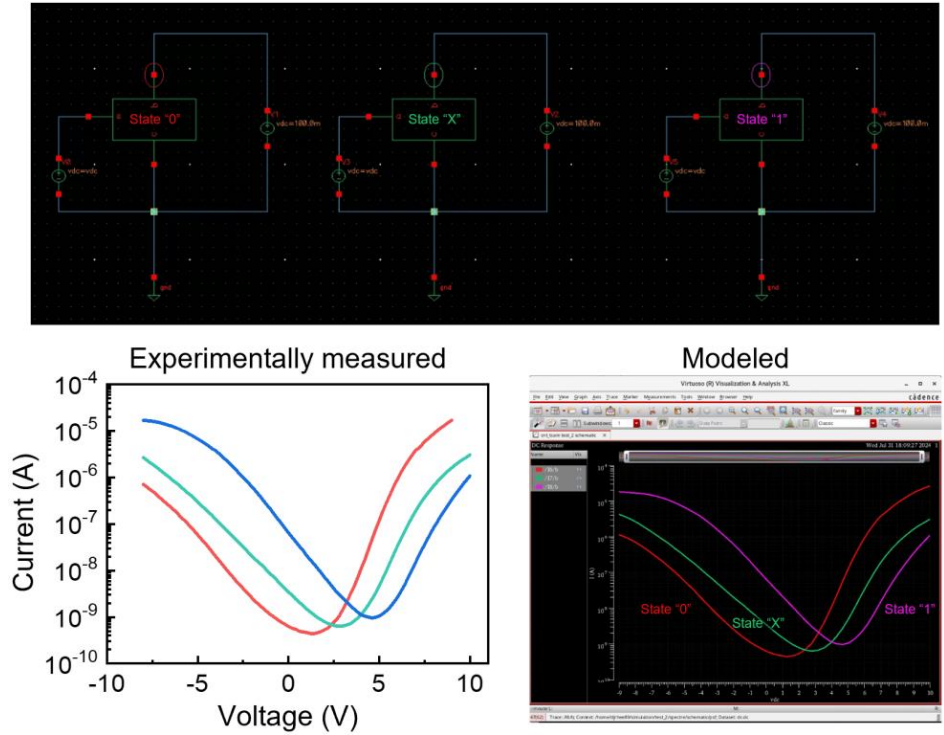

**Figure S10. Description of experimental transfer curves of the SWCNT FeFET in Cadence.**

*C* of pre-charge capacitor: 1 pF
$V_{SL}$ for search "1" : 6 V
$V_{SL}$ for search "0" : -2.6 V

**Figure S11. TCAM circuit diagram and simulated ML voltages for different stored states and $V_{SL}$.**

**Supplemenary information references**